\documentclass[aps,prb,twocolumn,superscriptaddress]{revtex4-2}
\usepackage{amsmath,amssymb,bm,graphicx}
\usepackage[T1]{fontenc}
\usepackage{hyperref}

\begin{document}

\title{Thermodynamic evidence for a pressure-driven crossover from strong- to weak-coupling superconductivity in Pb}

\author{Rustem Khasanov}
 \email{rustem.khasanov@psi.ch}
 \affiliation{PSI Center for Neutron and Muon Sciences CNM, 5232 Villigen PSI, Switzerland}

\date{\today}

\begin{abstract}
The thermodynamic critical field $B_{\rm c}$ provides direct access to the superconducting condensation energy, yet its pressure dependence has been studied much less extensively than that of the transition temperature. Here, muon-spin-rotation/relaxation measurements of the thermodynamic critical field $B_{\rm c}$ of elemental Pb under hydrostatic pressure up to $\simeq2.3$~GPa are reported. From the magnetic-field distribution in the intermediate state, $B_{\rm c}(T)$ is determined and $B_{\rm c}(0)$ is extracted at different pressures. In combination with previously reported high-pressure data for $B_{\rm c}$ and $T_{\rm c}$, it is shown that the pressure dependence of $B_{\rm c}(0)$ follows that of the superconducting gap $\Delta(0)$ more closely than that of the transition temperature $T_{\rm c}$. At higher pressures, the logarithmic pressure derivatives of $B_{\rm c}(0)$ and $T_{\rm c}$ are found to converge, indicating that the coupling strengths ratio $\alpha=\Delta(0)/k_{\rm B}T_{\rm c}$ becomes nearly pressure independent. This behavior is interpreted as thermodynamic evidence for a pressure-driven crossover from strong- to weak-coupling superconductivity in Pb.
\end{abstract}
\maketitle


\noindent {\it \underline{Introduction.}} Understanding how the characteristic energy scales of a superconductor evolve under external pressure provides important insight into the microscopic mechanisms governing superconductivity \cite{Lorenz_Chu_2005,Hamlin_PhysicaC_2015,Schilling_book_2007,Schilling_JPCS_2008}. In conventional phonon-mediated superconductors, pressure modifies both the electronic structure and the lattice dynamics, typically leading to phonon hardening and a reduction of the electron--phonon coupling strength \cite{Schilling_book_2007,Schilling_JPCS_2008,McMillan_PR_1968,Allen_Dynes_PRB_1975}. As a consequence, compression can drive the system gradually from the strong-coupling regime toward the weak-coupling Bardeen--Cooper--Schrieffer (BCS) limit \cite{Carbotte_RMP_1990}. Experimental signatures of such a crossover are usually inferred from the pressure dependence of the superconducting transition temperature $T_{\rm c}$. However, $T_{\rm c}$ is determined by the linearized gap equation and reflects a delicate balance between competing effects, including phonon hardening and changes in the electron--phonon interaction \cite{Eliashberg_JETP_1960,McMillan_PR_1968,Allen_Dynes_PRB_1975}. Thermodynamic quantities that probe the condensation energy can therefore provide a more direct view of the superconducting energy scale \cite{Carbotte_RMP_1990}.

One such quantity is the thermodynamic critical field $B_{\rm c}$, which is directly related to the free-energy difference between the superconducting and normal states \cite{Tinkham_1996,deGennes_1999}. The corresponding superconducting condensation-energy density is given by
\begin{equation}
U_0=\frac{B_{\rm c}^2(0)}{2\mu_0},
\label{eq:Free-Energy_Bc}
\end{equation}
where $B_{\rm c}(0)$ is the zero-temperature thermodynamic critical field. Within weak-coupling BCS theory, the condensation energy can also be written as \cite{Tinkham_1996,deGennes_1999,BCS_PR_1957,Schrieffer_1964}
\begin{equation}
U_0=\frac{1}{2}N(E_{\rm F})\Delta^2(0)
   =\frac{3}{4\pi^2 k_{\rm B}^2}\gamma_{\rm e} \Delta^2(0),
\label{eq:Free-Energy_Gamma}
\end{equation}
where $N(E_{\rm F})$ is the electronic density of states at the Fermi level, $\Delta(0)$ is the superconducting energy gap at zero temperature, $k_{\rm B}$ is the Boltzmann constant, and $\gamma_{\rm e}=(2\pi^2/3)k_{\rm B}^2N(E_{\rm F})$ is the Sommerfeld coefficient. Combining Eqs.~(\ref{eq:Free-Energy_Bc}) and (\ref{eq:Free-Energy_Gamma}) yields the approximate scaling relation
\begin{equation}
B_{\rm c}(0)\propto \Delta(0)\sqrt{\gamma_{\rm e}},
\label{eq:Bc_Gamma}
\end{equation}
showing that the thermodynamic critical field is governed by both the superconducting gap and the electronic density of states.

\begin{figure*}[htb]
\centering
\includegraphics[width=1.0\linewidth]{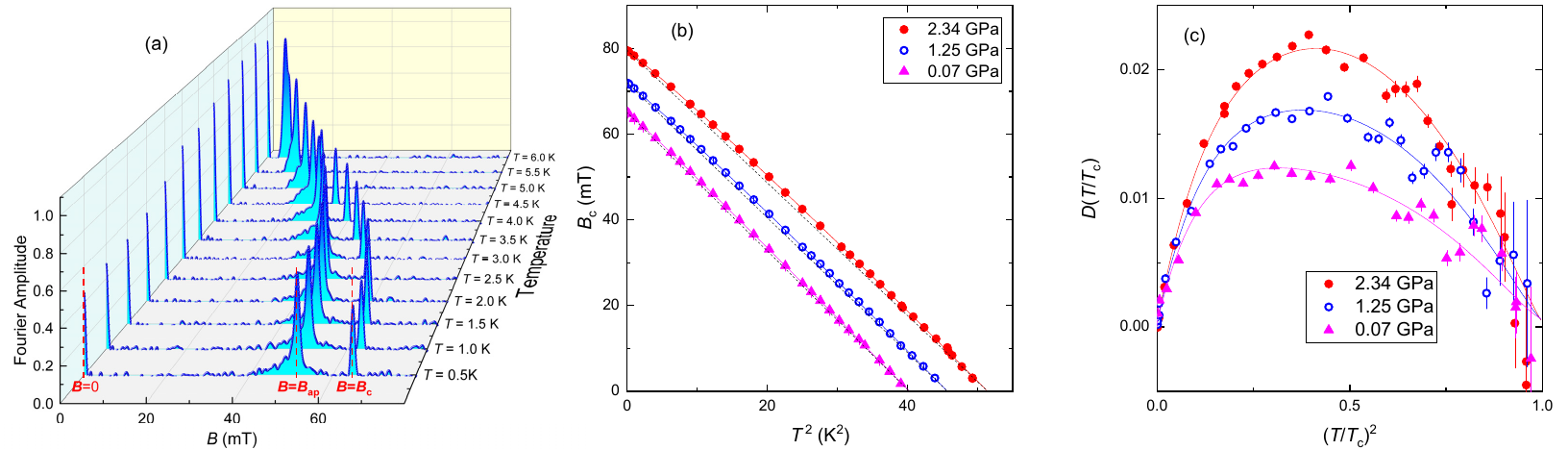}
\caption{
(a) Fourier transforms of the muon-time spectra at $p=2.34$~GPa, showing the magnetic-field distribution in the pressure-cell/sample assembly. Three distinct peaks are observed: (i) a peak at $B=0$, corresponding to muons stopping in superconducting domains of the Pb sample (Meissner state); (ii) a peak at the applied field, $B=B_{\rm ap}$, arising from muons stopping in the pressure-cell walls; and (iii) a peak at $B=B_{\rm c}$, corresponding to muons stopping in normal-state domains of the Pb sample.
(b) Temperature dependence of the thermodynamic critical field $B_{\rm c}(T)$ measured at pressures $p=0.07$, 1.25, and 2.34 GPa. The solid lines are fits within the $\alpha$ model, while the dashed lines represent parabolic dependences.
(c) Temperature dependence of the deviation function $D(T)=B_{\rm c}(T)-B_{\rm c}(0)[1-(T/T_{\rm c})^2]$, highlighting the deviation of $B_{\rm c}(T)$ from parabolic temperature dependence. The solid lines are fits using the $\alpha$ model.
}
\label{fig:Bc_Deviations}
\end{figure*}

It is convenient to express the gap in terms of the dimensionless ratio
\begin{equation}
\alpha=\frac{\Delta(0)}{k_{\rm B}T_{\rm c}},
\label{eq:alpha}
\end{equation}
where $T_{\rm c}$ is the superconducting transition temperature. The parameter $\alpha$ is commonly used as a measure of coupling strength via comparison with the weak-coupling BCS value $\alpha_{\rm BCS}\equiv1.764$. Values significantly above $\alpha_{\rm BCS}$ indicate enhanced strong-coupling effects, whereas values close to $\alpha_{\rm BCS}$ correspond to the weak-coupling limit. Combining Eqs.~(\ref{eq:Bc_Gamma}) and (\ref{eq:alpha}) gives the relation between the logarithmic pressure derivatives of $B_{\rm c}$, $T_{\rm c}$, $\gamma_{\rm e}$, and $\alpha$:
\begin{equation}
\frac{d\ln B_{\rm c}(0)}{dp} = \frac{d\ln T_{\rm c}}{dp} + \frac{1}{2}\frac{d\ln\gamma_{\rm e}}{dp} + \frac{d\ln\alpha}{dp}. \nonumber
\end{equation}
In simple metals, the pressure dependence of the electronic specific-heat coefficient $\gamma_{\rm e}$ is generally weak compared with that of $T_{\rm c}$ and $B_{\rm c}$ \cite{Hamlin_PhysicaC_2015,Schilling_book_2007,Schilling_JPCS_2008,Berman_JETP_1968}. Therefore, to a good approximation, the difference between the pressure derivatives of $B_{\rm c}$ and $T_{\rm c}$ reflects the pressure evolution of the gap ratio:
\begin{equation}
\frac{d\ln\alpha}{dp} \simeq \frac{d\ln B_{\rm c}(0)}{dp} - \frac{d\ln T_{\rm c}}{dp}.
 \label{eq:alpha_derivative}
\end{equation}
This relation shows that the pressure dependence of the thermodynamic critical field provides direct information on the pressure evolution of the gap ratio $\alpha$. If, in addition, $\alpha$ is found to decrease toward the weak-coupling BCS value $\alpha_{\rm BCS}\equiv 1.764$, the pressure evolution of $B_{\rm c}$ can be interpreted as evidence for a crossover from strong- to weak-coupling superconductivity.

In the present work, muon-spin rotation/relaxation ($\mu$SR) is used to determine the thermodynamic critical field of elemental Pb under hydrostatic pressure up to $\simeq2.3$~GPa. Unlike transport- or susceptibility-based determinations of field-dependent phase boundaries, $\mu$SR in the intermediate state provides direct access to the equilibrium thermodynamic critical field $B_{\rm c}$. Analysis of the temperature dependence $B_{\rm c}(T)$ within the $\alpha$ model \cite{Padamsee_JLTP_1973,Johnston_SST_2013} yields the pressure evolution of the key superconducting parameters $T_{\rm c}$, $B_{\rm c}(0)$, and $\Delta(0)$. It is shown that, in the investigated pressure range, the pressure dependence of $B_{\rm c}(0)$ follows that of the superconducting gap more closely than that of the transition temperature. By combining the present $\mu$SR results with previously reported high-pressure data \cite{Brandt_JETP_1975},  it is further found that the pressure derivatives of $B_{\rm c}(0)$ and $T_{\rm c}$ approach each other at higher pressures. Together with the experimentally observed decrease of $\alpha=\Delta/k_B T_c$ under pressure, this behavior is interpreted as thermodynamic evidence that compression drives Pb from the strong-coupling regime toward the weak-coupling limit.


\noindent {\it \underline{Experimental details.}} Details on sample preparation, transverse-field (TF) $\mu$SR experiments under pressure, the analysis of $\mu$SR data, and the $\alpha$ model used to describe the $B_{\rm c}(T,p)$ dependence are provided in the Supplemental Material.


\noindent {\it \underline{Experimental data.}} The TF-$\mu$SR measurements were carried out in the intermediate state, i.e., when the Pb sample separates into superconducting (Meissner) and normal-state domains \cite{deGennes_1999, Poole_Book_2014,Kittel_Book_1996,Prozorov_PRL_2007, Prozorov_NatPhys_2008, Khasanov_Bi-II_PRB_2019,Karl_PRB_2019,Khasanov_Ga-II_PRB_2020, Khasanov_AuBe_PRR_2020, Khasanov_PRB_2024, Khasanov_PRB_2025}. In this state, the magnetic induction in the superconducting domains is zero, whereas in the normal-state domains it is equal to the thermodynamic critical field, $B_{\rm c}$ \cite{Tinkham_1996,deGennes_1999}. This makes TF-$\mu$SR a particularly direct probe of $B_{\rm c}(T)$, since the field distribution measured by the implanted muons contains a distinct contribution from the normal domains at $B=B_{\rm c}$ \cite{Khasanov_Bi-II_PRB_2019,Karl_PRB_2019,Khasanov_Ga-II_PRB_2020, Khasanov_AuBe_PRR_2020,Khasanov_PRB_2024,Khasanov_PRB_2025,Egorov_PRB_2001,Leng_PRB_2019, Gladisch_HypInteract_1979,Grebinnik_JETP_1980,Beare_PRB_2019,Kozhevnikov_JSNM_2020}.

Representative Fourier transforms of the muon-time spectra, corresponding to the magnetic-field distribution in the pressure-cell/sample assembly at $p=2.34$~GPa, are shown in Fig.~\ref{fig:Bc_Deviations}(a). The field distribution consists of three well-defined peaks. The peaks at $B=0$ and $B=B_{\rm c}$ correspond to muons stopping in superconducting (Meissner-state) and normal-state domains of the Pb sample, respectively. The peak at the applied-field position, $B=B_{\rm ap}$, originates from muons stopping in the pressure-cell walls and precessing in the applied magnetic field. Such a three-peak field distribution is characteristic of TF-$\mu$SR measurements in the intermediate state of a type-I superconductor and directly illustrates how $B_{\rm c}$ is determined experimentally \cite{Khasanov_PRB_2024,Khasanov_PRB_2025,Egorov_PRB_2001,Leng_PRB_2019,Beare_PRB_2019,Khasanov_PRB_2021}.

The temperature dependence of the thermodynamic critical field obtained at $p=0.07$, 1.25, and 2.34 GPa is presented in Fig.~\ref{fig:Bc_Deviations}(b). The dashed lines show the parabolic dependence of $B_{\rm c}(T)$, which is used here as a reference. As discussed in Refs.~\onlinecite{Padamsee_JLTP_1973,Johnston_SST_2013}, the deviation of $B_{\rm c}(T)$ from this parabolic form is governed by the value of $\alpha=\Delta/(k_{\rm B}T_{\rm c})$. To illustrate this more clearly, Fig.~\ref{fig:Bc_Deviations}(c) shows the deviation function
$D(T)=B_{\rm c}(T)-B_{\rm c}(0)[1-(T/T_{\rm c})^2]$.
The solid lines in panels (b) and (c) are fits performed within the phenomenological $\alpha$ model, adapted for strong-coupling superconductors such as Pb \cite{Padamsee_JLTP_1973,Johnston_SST_2013,Khasanov_PRB_2021}. Within this model, the normalized gap function is taken in the BCS form, while $\alpha$ is treated as an adjustable parameter. The reduction of the deviation with increasing pressure therefore directly reflects the decrease of $\alpha$. Analysis of the $B_{\rm c}(T)$ data yields the pressure-dependent parameters $T_{\rm c}$, $B_{\rm c}(0)$, $\Delta(0)$, and $\alpha$ which form the basis for the pressure analysis presented below.

\begin{figure}[htb]
\centering
\includegraphics[width=0.85\linewidth]{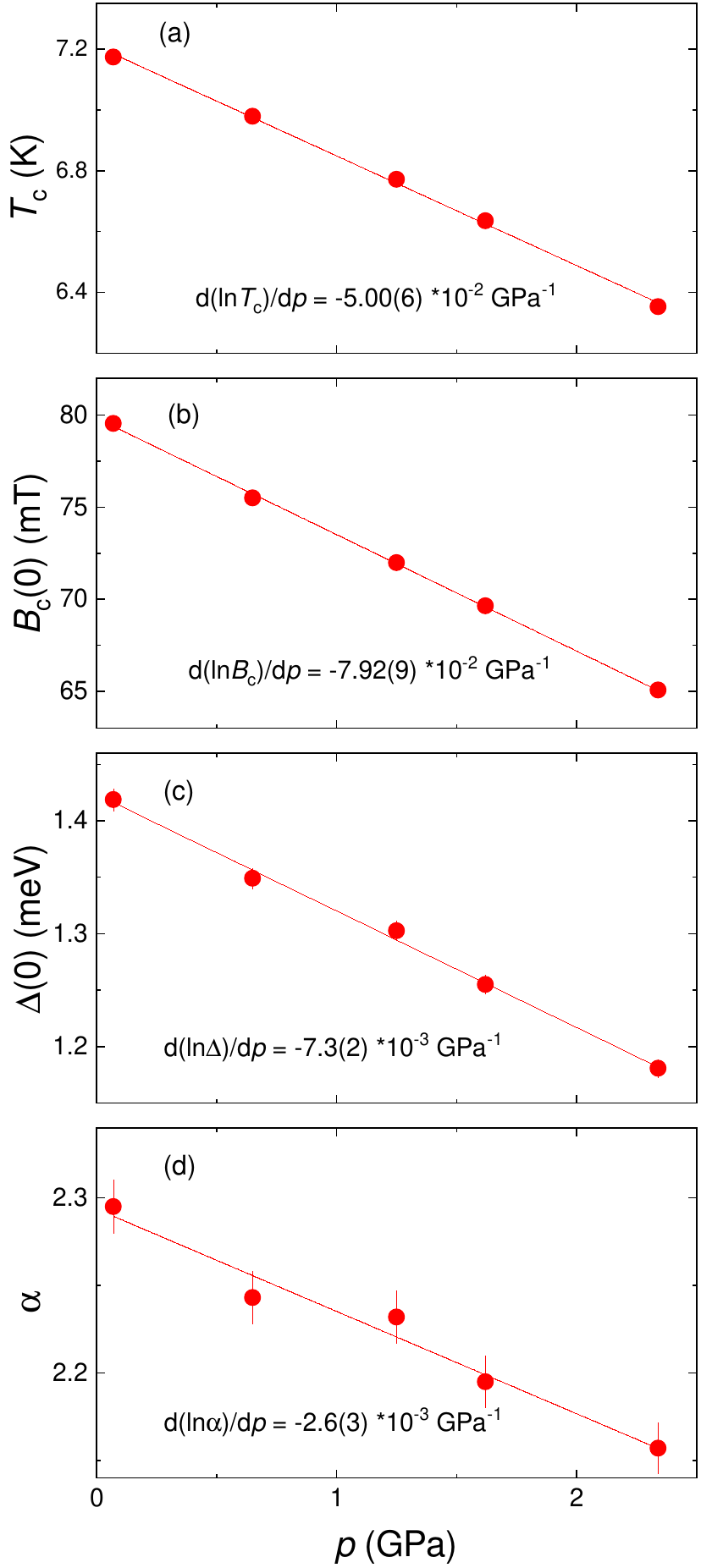}
\caption{
Pressure dependencies of the superconducting parameters obtained from fits of $B_c(T)$ data using the $\alpha$ model: (a) superconducting transition temperature $T_c(p)$, (b) thermodynamic critical field at zero temperature $B_c(0,p)$, (c) superconducting energy gap $\Delta()0,p)$, and (d) the coupling parameter $\alpha(p)$. Solid lines represent linear fits to the data. The corresponding logarithmic pressure derivatives are indicated.
}
 \label{fig:Parameters}
\end{figure}

Figure~\ref{fig:Parameters} presents the pressure dependences of the parameters extracted from the fits of $B_{\rm c}(T)$: $T_{\rm c}(p)$ [panel (a)], $B_{\rm c}(0,p)$ [panel (b)], $\Delta(0,p)$ [panel (c)], and $\alpha(p)$ [panel (d)]. All four quantities decrease approximately linearly over the investigated pressure range. The linear fits yield the logarithmic pressure derivatives ${\rm d}\ln T_{\rm c}/{\rm d}p=-5.00(6)\times10^{-2}$~GPa$^{-1}$, ${\rm d}\ln B_{\rm c}(0)/{\rm d}p=-7.92(9)\times10^{-2}$~GPa$^{-1}$, ${\rm d}\ln \Delta(0)/{\rm d}p=-7.3(3)\times10^{-2}$~GPa$^{-1}$, and ${\rm d}\ln \alpha/{\rm d}p=-2.6(3)\times10^{-2}$~GPa$^{-1}$. Remarkably, the logarithmic pressure coefficient of $\alpha$ is nearly equal to the difference between those of $B_{\rm c}(0)$ and $T_{\rm c}$, consistent with Eq.~(\ref{eq:alpha_derivative}).


\noindent {\it \underline{Discussion.}}The central idea of the paper is that the thermodynamic critical field provides access to the superconducting condensation energy and therefore probes the superconducting energy scale in a way that is complementary to the transition temperature alone. The present $\mu$SR measurements provide a direct realization of this concept for elemental Pb.

The low-pressure results summarized in Figs.~\ref{fig:Bc_Deviations} and \ref{fig:Parameters} establish a consistent picture. First, the temperature dependence $B_{\rm c}(T)$ is well described by the $\alpha$ model, and the deviation from a simple parabolic form is reduced systematically with increasing pressure. Second, the pressure dependences extracted from these fits show that $B_{\rm c}(0)$ and $\Delta(0)$ evolve in a very similar manner, whereas $T_{\rm c}$ decreases more slowly. Third, the pressure dependence of the gap ratio $\alpha=\Delta/(k_{\rm B}T_{\rm c})$ is consistent with the thermodynamic relation introduced in Eq.~(\ref{eq:alpha_derivative}).

These observations follow naturally from the framework developed above. Since in simple metals the pressure dependence of the Sommerfeld coefficient $\gamma_{\rm e}$ is relatively weak \cite{Hamlin_PhysicaC_2015,Schilling_book_2007,Schilling_JPCS_2008,Berman_JETP_1968}, the nearly equal logarithmic pressure derivatives of $B_{\rm c}(0)$ and $\Delta(0)$ reported in Figs.~\ref{fig:Parameters}~(b) and (c) represent the expected thermodynamic signature of a condensation energy governed primarily by the superconducting gap; see Eqs.~(\ref{eq:Free-Energy_Bc})--(\ref{eq:Bc_Gamma}). At the same time, according to Eq.~(\ref{eq:alpha_derivative}), the weaker pressure dependence of $T_{\rm c}$ implies that the gap ratio $\alpha$ must evolve with pressure. This is seen directly in Fig.~\ref{fig:Parameters}(d), where $\alpha$ decreases approximately linearly over the measured pressure range. Physically, this behavior is consistent with the known strong-coupling character of Pb at ambient pressure \cite{Carbotte_RMP_1990, Khasanov_PRB_2021, McMillan_Rowell_PRL_1965, Chainani_PRL_2000}: compression hardens the phonon spectrum and reduces the electron--phonon interaction, thereby driving the system toward weaker coupling.

\begin{figure}[htb]
\centering
\includegraphics[width=0.95\linewidth]{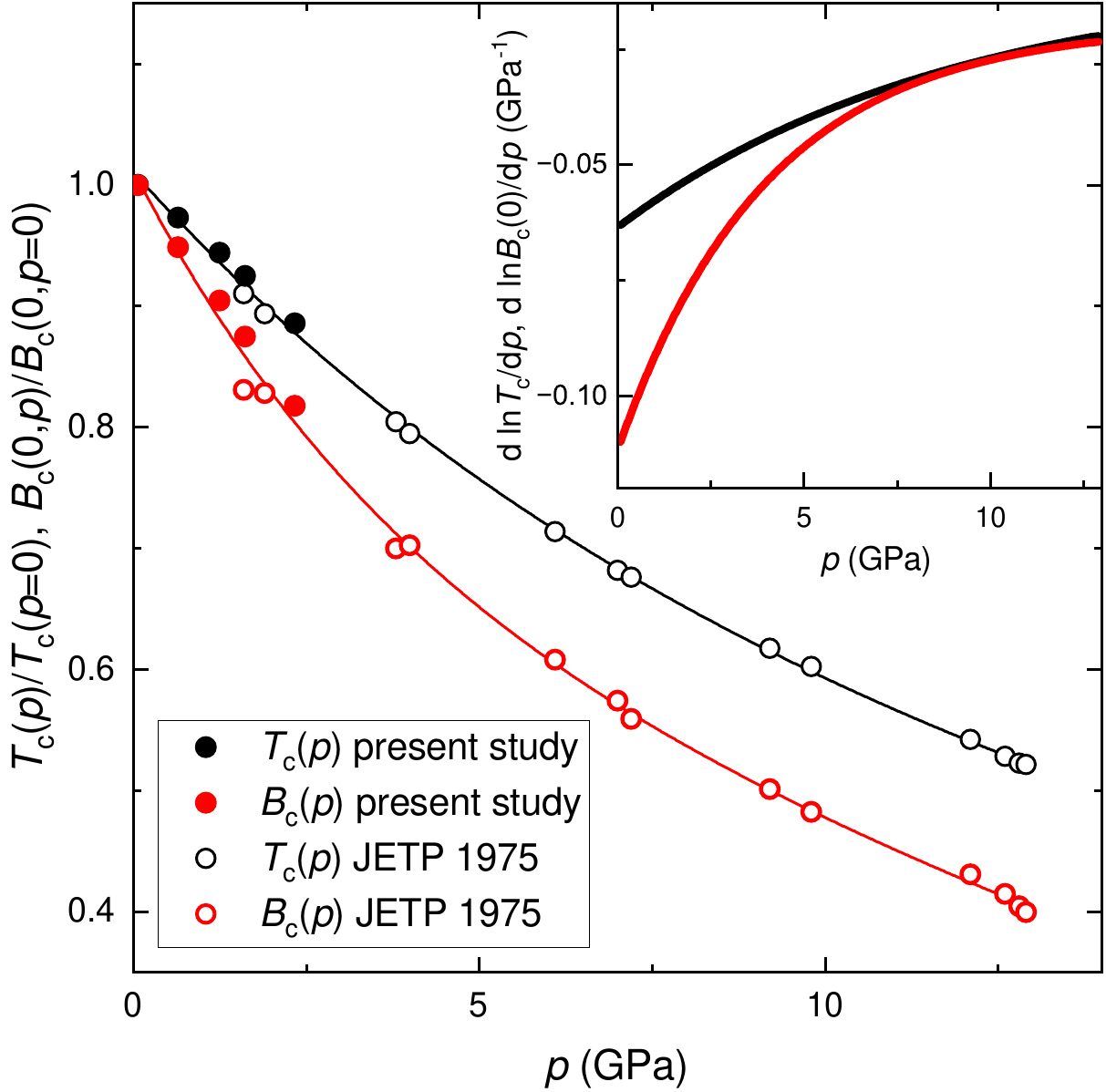}
\caption{
Normalized pressure dependences of the superconducting transition temperature $T_c(p)/T_c(p=0)$ and the thermodynamic critical field $B_c(0,p)/B_c(0, p=0)$ for Pb. Closed symbols represent the present $\mu$SR data, while open symbols correspond to literature data from Brandt \textit{et al.} Ref.~\onlinecite{Brandt_JETP_1975}. The solid lines are empirical fits using a double-exponential form.
Inset: Pressure dependences of the logarithmic derivatives $d\;\ln T_c/dp$ and $d\; \ln B_c(0)/dp$ obtained from the fit functions. The slopes approach each other and become nearly equal above $p\sim 8$ GPa, indicating convergence of the pressure responses of $T_c$ and $B_c$.
}
 \label{fig:Tc-Bc_vs_p}
\end{figure}

The present $\mu$SR experiment is restricted to the low-pressure regime, but it already establishes the relevant trend. To determine how this tendency develops at higher pressure, Fig.~\ref{fig:Tc-Bc_vs_p} compares the present data with the high-pressure results of Brandt \textit{et al.} \onlinecite{Brandt_JETP_1975} for $B_{\rm c}(0,p)$ and $T_{\rm c}(p)$. In the main panel, the normalized pressure dependences of $T_{\rm c}$ and $B_{\rm c}(0)$ differ most strongly at low pressure, where the pressure evolution of $\alpha$ is largest. With increasing pressure, however, the slopes of the two curves gradually approach each other. This becomes especially clear in the inset of Fig.~\ref{fig:Tc-Bc_vs_p}, which shows the logarithmic derivatives ${\rm d}(\ln T_{\rm c})/{\rm d}p$ and ${\rm d}(\ln B_{\rm c})/{\rm d}p$ obtained from the empirical fit functions.

The key result of this extended comparison is that the logarithmic pressure derivatives of $T_{\rm c}$ and $B_{\rm c}(0)$ become nearly identical for $p\gtrsim 8$~GPa, see the inset of Fig.~\ref{fig:Tc-Bc_vs_p}. Within Eq.~(\ref{eq:alpha_derivative}), this means that ${\rm d}\ln\alpha/{\rm d}p$ becomes very small in this pressure range. In other words, the strong-coupling corrections to the gap ratio cease to evolve significantly, and Pb approaches a regime in which $\alpha$ is nearly pressure independent. Since the low-pressure $\mu$SR data show that $\alpha$ decreases with increasing pressure, the subsequent convergence of ${\rm d}\ln B_{\rm c}(0)/{\rm d}p$ and ${\rm d}\ln T_{\rm c}/{\rm d}p$ at higher pressures is naturally interpreted as thermodynamic evidence that Pb evolves from the strong-coupling regime toward the weak-coupling limit.

More broadly, the present analysis clarifies why the thermodynamic critical field is a particularly informative quantity in pressure studies of conventional superconductors. Whereas $T_{\rm c}$ reflects the onset of superconductivity through the linearized gap equation, $B_{\rm c}$ probes the fully developed superconducting state through the condensation energy. The combined analysis of $B_{\rm c}(T)$, the extracted pressure dependences of $B_{\rm c}(0)$, $T_{\rm c}$, $\Delta(0)$, and $\alpha$, together with the comparison to higher-pressure data, therefore yields a coherent thermodynamic picture of how the superconducting energy scales of Pb evolve under compression.


\noindent {\it \underline{Conclusions.}} The thermodynamic critical field of elemental Pb under hydrostatic pressure up to $2.3$~GPa has been determined by means of muon-spin rotation/relaxation measurements in the intermediate state. Analysis of the resulting $B_{\rm c}(T)$ curves within the $\alpha$ model has provided the pressure dependences of the zero-temperature critical field $B_{\rm c}(0)$, the transition temperature $T_{\rm c}$, the superconducting energy gap $\Delta(0)$, and the gap ratio $\alpha$ in the low-pressure regime.

It has been found that, at low pressure, $B_{\rm c}(0)$ follows $\Delta(0)$ much more closely than it follows $T_{\rm c}$. This represents the central thermodynamic result of the present work and follows naturally from the relation between $B_{\rm c}$ and the condensation energy outlined above. At the same time, the weaker pressure dependence of $T_{\rm c}$ implies a decrease of the gap ratio $\alpha=\Delta(0)/(k_{\rm B}T_{\rm c})$, indicating that strong-coupling effects in Pb are already suppressed within the experimentally accessible pressure range.

By combining the present $\mu$SR results with previously reported high-pressure measurements, it has further been shown that the logarithmic pressure derivatives of $B_{\rm c}(0)$ and $T_{\rm c}$ gradually converge and become nearly equal above $p\sim 8$~GPa. Within the thermodynamic framework introduced in this work, this implies that the pressure evolution of $\alpha$ becomes very weak at high pressure. Together with the experimentally observed decrease of $\alpha$ at low pressure, this behavior is taken as thermodynamic evidence that compression drives Pb from the strong-coupling regime toward the weak-coupling limit.

More generally, it has been demonstrated that measurements of the thermodynamic critical field under pressure provide information that is complementary to that obtained from the transition temperature alone. Because $B_{\rm c}$ probes the condensation energy of the superconducting state, it provides direct access to the evolution of the superconducting energy scale and coupling strength under compression.

\vspace{0.5cm}
\noindent {\it \underline{Acknowledgments.}} The author acknowledges helpful discussion with Giovanni Ummarino.


\begin{thebibliography}{99}

\bibitem{Lorenz_Chu_2005} B. Lorenz and C. W. Chu, {\it High Pressure Effects on Superconductivity}, in {\it Frontiers in Superconducting Materials}, edited by A. V. Narlikar (Springer, Berlin, Heidelberg, 2005), pp. 459--497.\\
\url{https://doi.org/10.1007/3-540-27294-1_12}

\bibitem{Hamlin_PhysicaC_2015} J. J. Hamlin, {\it Superconductivity in the metallic elements at high pressures}, Physica C {\bf 514}, 59--76 (2015).\\
\url{https://doi.org/10.1016/j.physc.2015.02.032}

\bibitem{Schilling_book_2007} J.S. Schilling, {\it High-Pressure Effects}, in {\it Handbook of High-Temperature Superconductivity}, edited by J. R. Schrieffer and J. S. Brooks (Springer, New York, 2007). \\
    \url{https://doi.org/10.1007/978-0-387-68734-6_11}

\bibitem{Schilling_JPCS_2008} J.S. Schilling and J. J. Hamlin, {\it Recent studies in superconductivity at extreme pressures}, J. Phys.: Conf. Ser. {\bf 121}, 052006 (2008).\\
    \url{https://doi.org/10.1088/1742-6596/121/5/052006}

\bibitem{McMillan_PR_1968} W. L. McMillan, {\it Transition Temperature of Strong-Coupled Superconductors}, Phys. Rev. {\bf 167}, 331--344 (1968).\\
\url{https://doi.org/10.1103/PhysRev.167.331}

\bibitem{Allen_Dynes_PRB_1975} P. B. Allen and R. C. Dynes, {\it Transition temperature of strong-coupled superconductors reanalyzed}, Phys. Rev. B {\bf 12}, 905--922 (1975).\\
\url{https://doi.org/10.1103/PhysRevB.12.905}

\bibitem{Carbotte_RMP_1990} J. P. Carbotte, {\it Properties of boson-exchange superconductors}, Rev. Mod. Phys. {\bf 62}, 1027--1157 (1990).\\
\url{https://doi.org/10.1103/RevModPhys.62.1027}

\bibitem{Eliashberg_JETP_1960} G. M. Eliashberg, {\it Interactions between electrons and lattice vibrations in a superconductor}, Sov. Phys. JETP {\bf 11}, 696--702 (1960).\\
\url{https://www.jetp.ras.ru/cgi-bin/dn/e_011_03_0696.pdf}

\bibitem{Tinkham_1996} Michael Tinkham, {\it Introduction to Superconductivity}, 2nd ed. (McGraw-Hill, New York, 1996).\\
\url{https://search.worldcat.org/title/Introduction-to-superconductivity/oclc/32625121}

\bibitem{deGennes_1999} P. G. de Gennes, {\it Superconductivity of Metals and Alloys} (CRC Press, Boca Raton, 1999).\\
\url{https://www.routledge.com/Superconductivity-Of-Metals-And-Alloys/DeGennes/p/book/9780738201016}

\bibitem{BCS_PR_1957} J. Bardeen, L. N. Cooper, and J. R. Schrieffer, {\it Microscopic Theory of Superconductivity},  Phys. Rev. {\bf 106}, 162 (1957).\\
    \url{https://doi.org/10.1103/PhysRev.106.162}

\bibitem{Schrieffer_1964} J. R. Schrieffer, {\it Theory of Superconductivity} (Addison-Wesley, Reading, Massachusetts, 1964).\\
\url{https://search.worldcat.org/title/theory-of-superconductivity/oclc/932305355}

\bibitem{Berman_JETP_1968} I. V. Berman, N. B. Brandt, and N. I. Ginzburg, {\it Investigation of the effect of pressures up to 30 katm on the critical field of tin and indium at temperatures of 0.1--4 K}, Sov. Phys. JETP {\bf 26}, 86--91 (1968).\\
\url{https://www.jetp.ras.ru/cgi-bin/dn/e_026_01_0086.pdf}

\bibitem{Padamsee_JLTP_1973} H. Padamsee, J. E. Neighbor, and C. A. Shiffman, {\it Quasiparticle phenomenology for thermodynamics of strong-coupling superconductors},  J. Low Temp. Phys. {\bf 12}, 387 (1973).\\
    \url{https://doi.org/10.1007/BF00654872}

\bibitem{Johnston_SST_2013} D.C. Johnston, {\it Elaboration of the $\alpha-$model derived from the BCS theory of superconductivity}, Supercond. Sci. Technol. {\bf 26}, 115011 (2013).\\
    \url{https://doi.org/10.1088/0953-2048/26/11/115011}

\bibitem{Brandt_JETP_1975} N. B. Brandt, I. V. Berman, and Yu. P. Kurkin, {\it Investigation of the critical field curves of lead at pressures up to 130 kbar and temperatures down to 0.1 K}, Sov. Phys. JETP {\bf 42}, 869--873 (1975) [Zh. Eksp. Teor. Fiz. {\bf 69}, 1710 (1975)].

\bibitem{Poole_Book_2014} C. P. Poole, Jr., H. A. Farach, R. J. Creswick, and R. Prozorov, {\it Superconductivity}, 3rd ed. (Academic Press, Amsterdam, 2014).\\
    \url{https://doi.org/10.1016/C2012-0-07073-1}

\bibitem{Kittel_Book_1996} C. Kittel, {\it Introduction to Solid State Physics}, 7th ed. (Wiley, New York, 1996).

\bibitem{Prozorov_PRL_2007} R. Prozorov, {\it Equilibrium Topology of the Intermediate State in Type-I Superconductors of Different Shapes}, Phys. Rev. Lett. {\bf 98}, 257001 (2007).\\
    \url{https://doi.org/10.1103/PhysRevLett.98.257001}

\bibitem{Prozorov_NatPhys_2008} R. Prozorov, A. F. Fidler, J. R. Hoberg, and P. C. Canfield, {\it Suprafroth in type-I superconductors}, Nat. Phys. {\bf 4}, 327 (2008).\\
    \url{https://doi.org/10.1038/nphys888}

\bibitem{Khasanov_Bi-II_PRB_2019} R. Khasanov, M. M. Radonji\'{c}, H. Luetkens, E. Morenzoni, G. Simutis, S. Sch\"{o}necker, W. H. Appelt, A. \"{O}stlin, L. Chioncel, and A. Amato, {\it Superconducting nature of the Bi-II phase of elemental bismuth}, Phys. Rev. B {\bf 99}, 174506 (2019).\\
    \url{https://doi.org/10.1103/PhysRevB.99.174506}

\bibitem{Karl_PRB_2019} R. Karl, F. Burri, A. Amato, M. Doneg\`{a}, S. Gvasaliya, H. Luetkens, E. Morenzoni, and R. Khasanov, {\it Muon spin rotation study of type-I superconductivity: Elemental $\beta-$Sn}, Phys. Rev. B {\bf 99}, 184515 (2019).\\
    \url{https://doi.org/10.1103/PhysRevB.99.184515}

\bibitem{Khasanov_Ga-II_PRB_2020} R. Khasanov, H. Luetkens, A. Amato, and E. Morenzoni, {\it Structural phases of elemental Ga: Universal relations in conventional superconductors}, Phys. Rev. B {\bf 101}, 054504 (2020).\\
    \url{https://doi.org/10.1103/PhysRevB.101.054504}

\bibitem{Khasanov_AuBe_PRR_2020} R. Khasanov, R. Gupta, D. Das, A. Amon, A. Leithe-Jasper, and E. Svanidze, {\it Multiple-gap response of type-I noncentrosymmetric BeAu superconductor}, Phys. Rev. Research {\bf 2}, 023142 (2020).\\
    \url{https://doi.org/10.1103/PhysRevResearch.2.023142}

 \bibitem{Khasanov_PRB_2024} Rustem Khasanov and Giovanni A. Ummarino, {\it Pressure weakens coupling strength in In and Sn elemental superconductors}, Phys. Rev. B {\bf 110}, 214515 (2024).\\
\url{https://doi.org/10.1103/PhysRevB.110.214515}

\bibitem{Khasanov_PRB_2025} Rustem Khasanov, Riccardo Vocaturo, Oleg Janson, Andreas Koitzsch, Ritu Gupta, Debarchan Das, Nicola P. M. Casati, Maia G. Vergniory, Jeroen van den Brink, and Eteri Svanidze, {\it Influence of pressure on the properties of the multigap type-I superconductor BeAu}, Phys. Rev. B {\bf 111}, 104507 (2025).\\
\url{https://doi.org/10.1103/PhysRevB.111.104507}

\bibitem{Egorov_PRB_2001} V. S. Egorov, G. Solt, C. Baines, D. Herlach, and U. Zimmermann, {\it Superconducting intermediate state of white tin studied by muon-spin-rotation spectroscopy}, Phys. Rev. B {\bf 64}, 024524 (2001).\\
    \url{https://doi.org/10.1103/PhysRevB.64.024524}

\bibitem{Leng_PRB_2019} H. Leng, J.-C. Orain, A. Amato, Y. K. Huang, and A. de Visser, {\it Type-I superconductivity in the Dirac semimetal PdTe$_2$ probed by $\mu$SR}, Phys. Rev. B {\bf 100}, 224501 (2019).\\
    \url{https://doi.org/10.1103/PhysRevB.100.224501}

\bibitem{Gladisch_HypInteract_1979} M. Gladisch, D. Herlach, H. Metz, H. Orth, G. zu Putlitz, A. Seeger, H. Teichler, W. Wahl, and W. Wigand, {\it Muon spin rotation in superconductors}, Hyperfine Interact. {\bf 6}, 109 (1979).\\
    \url{https://doi.org/10.1007/BF01028778}

\bibitem{Grebinnik_JETP_1980} V. G. Grebinnik, I. I. Gurevich, V. A. Zhukov, A. I. Klimov, L. A. Levina, V. N. Maiorov, A. P. Manych, E. V. Mel'nikov, B. A. Nikol'skii, A. V. Pirogov, A. N. Ponomarev, V. S. Roganov, V. I. Selivanov, and V. A. Suetin, {\it Investigation of superconductors by the muon technique}, Zh. Eksp. Teor. Fiz. {\bf 79}, 518 (1980); [Sov. Phys. JETP {\bf 52}, 261 (1980)].

\bibitem{Beare_PRB_2019} J. Beare, M. Nugent, M. N. Wilson, Y. Cai, T. J. S. Munsie, A. Amon, A. Leithe-Jasper, Z. Gong, S. L. Guo, Z. Guguchia, Y. Grin, Y. J. Uemura, E. Svanidze, and G. M. Luke, {\it $\mu$SR and magnetometry study of the type-I superconductor BeAu}, Phys. Rev. B {\bf 99}, 134510 (2019).\\
    \url{https://doi.org/10.1103/PhysRevB.99.134510}

\bibitem{Kozhevnikov_JSNM_2020} V. Kozhevnikov, A. Suter, T. Prokscha, and C. Van Haesendonck, {\it Experimental Study of the Magnetic Field Distribution and Shape of Domains Near the Surface of a Type-I Superconductor in the Intermediate State},  J. Supercond. Nov. Magn. {\bf 33}, 3361 (2020).\\
    \url{https://doi.org/10.1007/s10948-020-05576-1}

\bibitem{Khasanov_PRB_2021} Rustem Khasanov, Debarchan Das, Dariusz Jakub Gawryluk, Ritu Gupta, and Charles Mielke, {\it Isotropic single-gap superconductivity of elemental Pb}, Phys. Rev. B {\bf 104}, L100508 (2021).\\
\url{https://doi.org/10.1103/PhysRevB.104.L100508}

\bibitem{McMillan_Rowell_PRL_1965} W. L. McMillan and J. M. Rowell, {\it Lead Phonon Spectrum Calculated from Superconducting Density of States}, Phys. Rev. Lett. {\bf 14}, 108 (1965).\\
\url{https://doi.org/10.1103/PhysRevLett.14.108}

\bibitem{Chainani_PRL_2000} A. Chainani, T. Yokoya, T. Kiss, and S. Shin, {\it Photoemission Spectroscopy of the Strong-Coupling Superconducting Transitions in Lead and Niobium}, Phys. Rev. Lett. {\bf 85}, 1966 (2000).\\
\url{https://doi.org/10.1103/PhysRevLett.85.1966}



\end{thebibliography}
\end{document}